\documentclass[11pt]{article}
\usepackage{moriond2000,epsfig}

\begin{document}
\vspace*{4cm}
\title{ SPECTROSCOPIC FOLLOW UP OF ARCLETS IN AC114 WITH
THE VLT }

\author{ J.F. LE BORGNE$^1$, L. CAMPUSANO$^2$, R. PELLO$^1$,
J.P.KNEIB$^1$, R. ELLIS$^3$,
B. FORT$^4$, Y. MELLIER$^4$, I. SMAIL$^5$ }

\address{
$^1$ Laboratoire d'Astrophysique, OMP, Toulouse, France \\
$^2$ Observatorio Cerro Cal\'an, Dept. de Astronom\' \i a,
U. de Chile, Santiago, Chile \\
$^3$ Caltech, Passadena, USA  \\
$^4$ Institut d'Astrophysique de Paris, France \\
$^5$ Pysics Dept.,Durham University, Durham, UK \\
}

\maketitle\abstracts{ We present here the first results on the
VLT/FORS-1 spectroscopic survey of amplified sources and multiple images
in the lensing clusters AC114 and Cl0054.  Background sources were
selected in the cluster core, close to the critical lines, using
photometric redshifts combined with lensing inversion criteria.
Spectroscopic results are given, together with a brief summary of the
properties of some of these high-z galaxies.  }

\section{ Introduction }
 To study distant galaxies properties, we can either observe {\it blank
fields} or use clusters of galaxies as natural telescopes.  Cluster
lenses allow to study in greater details the same population seen in
blank field surveys and to probe intrinsically fainter galaxies that can
otherwise not be studied.  Of particular interest is to define a new
sample of high-z galaxies in the cluster cores where the amplification
is the highest (typically 1 to 3 magnitudes). For these galaxies, the
signal to noise ratio in the spectra and the detection fluxes are
improved beyond the limits of conventional techniques.

Such lensing techniques have been succesfully applied in various
wavebands from UV to Submm. To get rid of the cluster galaxies,
the selection criteria of the distant galaxy candidates are
based on a combination of photometric redshits and lensing inversion
methods (Kneib et al. 1996; Ebbels et al. 1998, Pell\'o et al 1999a,b).  In our study,
photometric redshifts are computed from broad-band photometry on a
large wavelength interval. This allows to reduce errors on the
redshift determination, and reduces the biases towards or against a
particular type of object or redshift domain. Furthermore, it helps to
optimize the instrument choice for the spectroscopic follow-up
(visible vs. near-IR). Photometric redshifts were computed
through a standard minimization procedure, using the public code
{\it hyperz} (Bolzonella et al. 2000). This procedure uses a template
library of spectra derived from the new Bruzual \& Charlot
evolutionary code (GISSEL98, Bruzual \& Charlot 1993).

We report here   the results obtained using the
VLT/FORS instrument in the cluster-lenses AC114 (z=0.312).
More details will be given in a forthcoming paper
(Campusano et al., 2000)

\section{ Observations and data reduction}
The spectra were obtained with the multi-objet spectroscopy mode of
FORS1 at the UT1 of VLT on 1999 October 5. The grism used was G300V,
with a wavelength coverage between $\sim$4000\AA\ and $\sim$8600\AA,
and a wavelength resolution of $R=500$ for the 1" slit width
used.  3 masks were cut for AC114, with total exposure times of 2h15,
1h30 and 1h10 respectively. Spectra of the spectrophotometric standard
star Feige 110 were obtained for spectrophotometric calibration. The
data reduction was done with standard $IRAF$ packages. From the 3
masks, 61 spectra of objects were extracted and calibrated, some of
them being on several masks. Among these 61 objects, 30 are cluster
galaxies, 20 are background galaxies and 11 are foreground objects.

\section{Redshifts of amplified sources}

\begin{figure}[t]
\psfig{figure=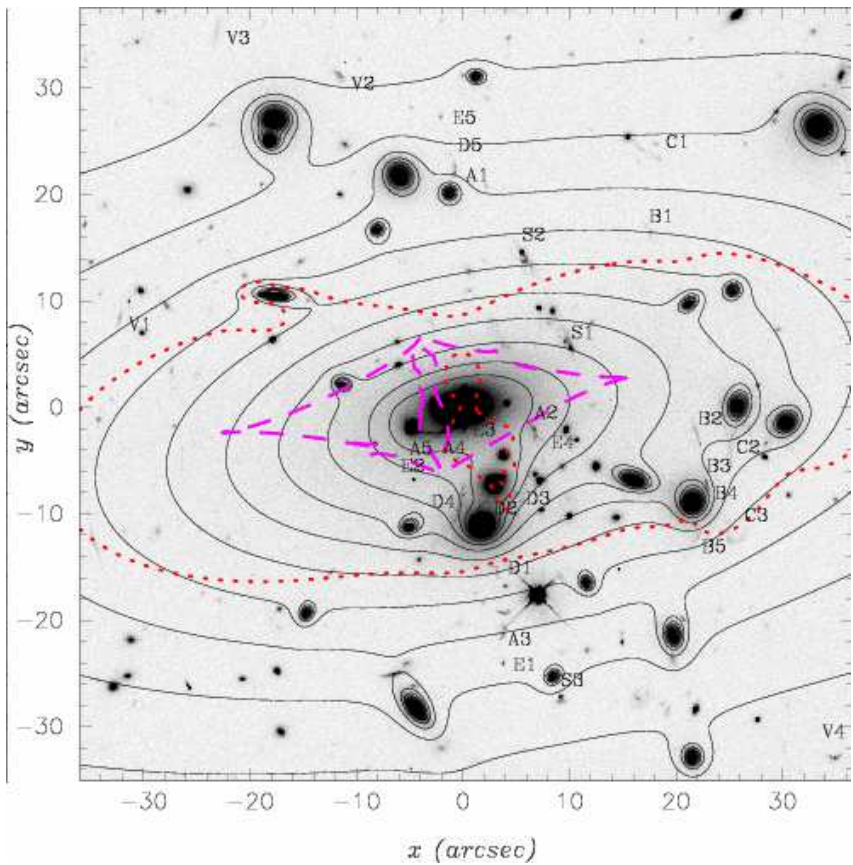, width=0.7\textwidth}
\caption{ \label{fig1}
HST/WFPC2 image (F702W) of AC114 cluster core with the identified high
redshift galaxies and multiple images. The thin contour lines
represent the total mass distribution as modelled by Natarajan et
al. (1998). The critical line (dotted lines) and the caustic lines
(dashed lines) are shown for $z=3.35$.  }
\end{figure}

The photometric selection in the cluster core was done using the
available data set, which includes deep  F702W/WFPC2 HST image,
J and K images obtained with SOFI at NTT and older U (ESO-NTT) and
V (Danish 1.5m telescope at ESO). In the present case, the lack of deep B images prevents a
selection in terms of pure photometric redshift criteria, in such a
way that the weight of the lens inversion method has been used
advantageously in the selection.
The size of the CCD images being different from one to another, the
photometric selection could only be done effectively on the central
HST field of $\sim$ 80''$\times$80''. This field is also ideal for
lens inversion thanks to the high resolution of HST images (Kneib et
al. 1996). As a consequence, all the objects selected for
spectroscopic follow up in the HST field, using both photometry and
lens inversion, are high-z candidates, and they were found to have
redshifts between $\sim 1$ and 3.5.  Our selection method was therefore
100\% efficient to identify high redshift galaxies in the cluster
core.
Figure \ref{fig1} displays the HST image of the central region of
AC114 in F702W, together with the identification
of the background sources. Table \ref{tab1} summarizes the information
about the redshifts and photometry for these objects. Magnitudes are
computed within arbitrary apertures, adapted to the shape of each
object, and fixed for all the filters.  The idea is to keep the same
physical region in the different filters when computing colors. Even
though photometric redshifts are highly uncertain in this case,
because of the lack of deep B photometry, we give the values found for
comparison, with the corresponding $1 \sigma$ error bars. In all cases
the photometric estimate is fully compatible with the spectroscopic
redshift within the errors.

We have computed the absolute M$_B$ magnitudes for these amplified
sources through a direct scaling of the observed SED, taking into
account the spectroscopic z, and using the best-fit templates from the
Bruzual \& Charlot code (Bruzual \& Charlot 1993) to derive the
k-corrections. The values, corrected for amplification, given in Table
\ref{tab1}, are computed using an estimated value of the amplification
obtained through
the Natarayan et al. (1998) model. These results, ranging from M$_B$
$\sim -18$ to $-20$, show that these objects are substantially fainter
than the corresponding ones in the present conventional studies.

\begin{table}[t]
\label{fig4}
\begin{center}
\begin{tabular}{lp{2.5in}}
\psfig{figure=fig2.ps,angle=270,width=0.5\textwidth}
&
Figure \ref{fig4}:
Spectrum of object S2 which confirm the identification of
Smail et al 1995. Note the blue excess in that galaxy.
\\
\end{tabular}
\end{center}
\end{table}

\subsection{The multiple image S at z=1.867}

The previous determination of the redshift by Smail et al. (1995) is
confirmed by the present results. Figure \ref{fig4} displays the high
S/N spectrum as well as the line identification. This line identification
is not complete, more work has to be done.

\subsection{The multiple image system E at z=3.347}

The spectrum of this compact source shows
a strong Ly$_\alpha$ in emission, in good agreement with the
photometric estimate. From the morphologic point of view, it is
compact on the WFPC images with a faint extension towards the NW
direction. Because of its location in the cluster core, it is found to
be a multiple-image showing a clear 5 image lensing configuration. The
other 4 images are also identified on Figure \ref{fig1}. Given its
morphology and the width of the emission line, this galaxy might be a
seyfert 2.

\begin{table}[t]
\caption{ \label{tab1} Characteristics of some background sources
 studied in the core of AC114. When available, the identifications are
 given according to Natarayan et al. 1998. $M_B$ have been corrected
 for magnification. Error bars in
 photometric redshifts correspond to $1 \sigma$.  }
\vspace{0.4cm}
\begin{center}
\begin{tabular}{|l|l|l|l|l|l|l|}
\hline
& & & & & & \\
 Id.   & $z$      & $z$       & $\Delta$z & R     & $M_B$  & $M_B$  \\
       &(spectro)&(photometric)&          &       &        & (corrected)\\
& & & & & & \\
\hline
A1-A2  &  1.605   & 1.00/1.69 & 0.68-1.91 & 23.51 & -21.89 & $\sim$-20. \\
B2-B3  &  1.66:   & 1.58      & 1.20-1.92 & 24.01 & -22.74 & $\sim$-20. \\
C3     &  2.84:   & 2.56      & 2.44-2.84 & 22.50:& -22.50 & -20.0 \\
E1     &  3.34695 & 3.15      & 2.46-3.43 & 24.18 & -21.79 & -20.4 \\
S2     &  1.86710 & 2.25      & 1.66-2.57 & 21.90 & -22.57 & -20.5 \\
V2     &  1.2143  & 1.53      & 1.19-1.75 & 23.22 & -20.80 & -20.2 \\
V3     &  2.08    & 2.29      & 0.95-2.74 & 24.36 & -22.36 &  -21.7  \\
V4     &  2.90    & 2.71      & 2.18-2.80 & 23.15 & -22.93 &  -22.3  \\%
& & & & & & \\
\hline
\end{tabular}
\end{center}
\end{table}


\end{document}